\documentclass[11pt]{article}

\usepackage{amsmath, amssymb, amsthm, mathtools}
\usepackage{amsmath,amssymb,amsthm,amsfonts}
\usepackage{dsfont}
\usepackage{bbm}
\usepackage[nice]{nicefrac}
\usepackage{graphicx}      
\usepackage{float}         
\usepackage{booktabs}      
\usepackage{multirow}      
\usepackage{array}         
\usepackage{float}

\usepackage{subfig}

\usepackage{url}           
\usepackage{hyperref}      
\hypersetup{
    hyperfootnotes=false,
    colorlinks=true,        
	linkcolor=black,    
    citecolor=black,         
    pdfborder={0 0 0}       
}
\usepackage[numbers,sort&compress]{natbib} 
\usepackage{bibentry}      

\usepackage{enumitem}      

\usepackage[letterpaper]{geometry} 
\newcommand{\G}{\mathcal{G}}

\title{Tripartite models for estimating the value of drug candidates and decision tools}

\author{
    J. Mellnik\thanks{Leigham Strategy, Santa Cruz, CA, USA. \texttt{jm.leigham@gmail.com}}  \and
    J. W. Scannell\thanks{JW Scannell Analytics Ltd., Edinburgh, UK \& Etheros Pharmaceuticals Corp., Nashville, TN, USA} 
}

\begin{document}

\maketitle

    \begin{abstract}
        Consider two similar drug companies with access to similar chemical libraries and synthesis methods, who each run an R\&D program. The programs have the same number of stages, which each take the same amount of time, with the same costs, with the same historic stepwise progression rates, and which aim to address the same therapeutic indication. Now let us suppose one of these companies invests in new scientific tools that make it unusually good at critical progression decisions, while the other company does not. How do we assess the difference in value between the two programs? Surprisingly, standard discounted cash flow valuation methods, such as risk-adjusted net present value (rNPV), ubiquitous in drug industry portfolio management and venture capital, are largely useless in this case. They fail to value the decisions that make drug candidates more or less valuable because rNPV conflates wrong decisions to progress bad candidates with right decisions to progress good ones. The purpose of this paper is to set out a new class of valuation model that logically links the value of therapeutic assets with the value of ``decisions tools” that are used to design, optimize, and test those assets. Our model makes clear the interaction between asset value and decision tool value.  It also makes clear the downstream consequences of better, or worse, upstream decisions. This new approach may support more effective allocation of R\&D capital; helping fund therapeutic assets that are developed using good decision tools, and funding better decision tools to distinguish between good and bad therapeutic assets.  
    \end{abstract}


\section{Critiques of rNPV}
\label{sec:rNPV_Critiques}

The canonical approach to valuing a drug in development is to compute its rNPV. Specifically, for a drug candidate with $N$ development stages before regulatory approval, let $c_{n}$ represent the cost of completing stage $n$ and $p_{n}$ represent the probability that the candidate progresses from stage $n$ to stage $n+1$. This is often called the ``probability of success” but it more accurately refers to stepwise progression probability. If $r$ is the reward received following regulatory approval (all costs and rewards discounted appropriately), then rNPV is

\begin{align}
    rNPV=rp_1p_2\dots p_N-\left(c_1p_0+c_2p_0p_1+\ldots c_Np_0p_1\ldots p_{N-1}\right) \label{eq:rNPV}
\end{align}

\noindent The first term in Eq. \ref{eq:rNPV} is the risk-adjusted reward, i.e. the size of the reward multiplied by the probability it will be achieved, and the second term is the risk-adjusted cost of development, i.e. the sum of the cost of each stage of development weighted by the probability that the drug candidate will advance to that stage.

\vspace{0.2cm}

The canonical approach has been the subject of a range of criticisms \cite{Rea2021, Walker2015, Rogers2002, Faulkner1996}. Here, we highlight two problems with conventional rNPV models. 

\vspace{0.2cm}

First, rNPV is a point estimate because it uses a single number to convey value. While point estimates may be useful for making quick comparisons, they fail to capture the range of potential outcomes and, as a result, don’t accurately convey the risk associated with a drug development program despite the fact that some adjustment for risk has occurred. To more accurately gauge risk, models that result in probability distributions over a range of possible outcomes should be used \cite{Walker2015}. Monte Carlo simulations can be helpful in estimating risk, however technical challenges make it difficult to accurately estimate the probabilities of rare events, which are often of the greatest interest (e.g. developing a blockbuster drug)  \cite{Biondini2015}. 

\vspace{0.2cm}

Second, any conventional rNPV-based approach is limited by a fundamental insensitivity to decision quality at each step in the R\&D process.

\vspace{0.2cm}

Decision-theoretic and historical analyses of drug R\&D show that productivity is very sensitive to the \emph{predictive validity} of the \emph{decision tools} that are used to distinguish between the therapeutic candidates that progress to the next stage and those that are terminated \cite{Scannell2016,Scannell2022}. Predictive validity is the extent to which a candidate’s performance, as measured by the decision tool, corresponds with its ``true" capacity for commercial and technical success. The term ``decision tool” encompasses any means for evaluating drug candidates for the purpose of determining which are worth continuing to develop and which are not. Decision tools include everything from a technology platform such as an AI model for binding prediction, to an in vitro assay, or even a clinical trial design \cite{Scannell2022}.

\vspace{0.2cm}

Conventional rNPV models fail to capture decision quality because $p_n$, the probability of progression, conflates incorrect decisions to progress bad candidates with correct decisions to progress good candidates \cite{Scannell2016, Ewart2022}. Furthermore, they fail to capture the downstream consequences of those upstream decisions because there is no logical link between $p_n$ and $p_{n+1}$ \cite{Scannell2016, Ewart2022}. 

\vspace{0.2cm}

Consider, for example, two drug candidates, A and B, both already known to be safe and well tolerated, that are evaluated in two Phase 2 clinical trials that have the same cost, duration, and recruit from the same patient population. The Drug A study uses conventional endpoints for the therapy area which are known to have a low signal to noise ratio (e.g., from high placebo response rates). The Drug B study, on the other hand, measures a new biomarker that is causally related to the pathophysiology in question, shows no placebo effect, and has a high signal to noise ratio. 

\vspace{0.2cm}

The probability of progression to Phase 3 from the two studies may be equal, but the value of the information gained from the studies is different, as are the downstream consequences of the studies. In the case of Drug A, our intuition is that the value of the drug candidate is barely changed because our understanding of whether it is likely to work barely changes (the trial had a low signal to noise ratio), whereas in the case of Drug B (with the high signal to noise ratio), our intuition is that the value of the program has increased more than that of Drug A. Standard rNPV models fail to reflect this intuition. It could, of course, be argued in this instance that the probabilities of success of Drug B’s Phase 3 trial could be revised upwards versus the historic norms. However, the rNPV framework provides no quantitative guidance on the size of the probability adjustment and hence the financial value of Drug B’s superior Phase 2 decision tool. A more effective approach would be a valuation framework that can simultaneously value therapeutics candidates, decision tools, and the interactions between the two. 

\section{A Tripartite Model}
\label{sec:SingleTripartite}

        Such simultaneous valuation can be achieved via what we refer to as a ``tripartite model". The model is described as tripartite because it contains information at three distinct levels: the observed result of a decision tool ($\hat{\delta}$), the unobserved ground truth result of a decision tool ($\delta$), and the ground truth capacity of the drug candidate to achieve commercial and technical success ($\G$). The parameter $\G$ is needed because regulatory approval does not guarantee commercial and technical success \cite{Dyer2024} and it is treated as a continuous variable because commercial and technical success is not a binary outcome \cite{Schuhmacher2023}. $\hat{\delta}$ is an estimate of $\delta$ and is typically assumed to be normally distributed with mean $\delta$ and standard deviation $\sigma_{\hat{\delta}}$ , i.e. $\hat{\delta}\sim\mathcal{N}(\delta,\sigma_{\hat{\delta}})$. The relationship between $\delta$ and $\G$ is explicitly modeled and the extent to which $\delta$ predicts $\G$ corresponds to the predictive validity of the decision tool. If, for example, $\delta$ and $\G$ are jointly distributed according to a bivariate normal distribution with mean $\mu=\big(\begin{smallmatrix}
  \mu_{\G}\\
  \mu_{\delta}
\end{smallmatrix}\big)$ and covariance $\Sigma=\big(\begin{smallmatrix}
  \sigma^2_{\G} & \rho\sigma_{\G}\sigma_{\delta}\\
  \rho\sigma_{\G}\sigma_{\delta} & \sigma^2_{\delta}
\end{smallmatrix}\big)$, then the correlation, $\rho$, captures the predictive validity of the decision tool. 

\vspace{0.2cm}

Elements of the tripartite model resemble the formalism in Scannell \& Bosley (2016)\cite{Scannell2016} and Miller et al. \cite{Miller2018}, but the new treatment gives a financial focus, generalizes to all stages of development and adds realism by treating success as a continuous variable.

\vspace{0.2cm}

Let $P_0(\delta,\G)$ denote the joint distribution of $\delta$ and $\G$ describing the belief about the probable values of $\delta$ and $\G$ prior to the start of a trial, and let $P_0(\delta)$ and $P_0(\G)$ denote the marginal distributions of $P_0(\delta,\G)$, i.e. the ``marginal priors”. If $P(\delta,\G)$ is the joint distribution of $\delta$ and $\G$ after a successful trial, the ``marginal posteriors” of $P(\delta,\G)$, are

\begin{align}
    P(\delta)=P(\text{Study success}|\delta)P_0(\delta) \label{eq:Pdelta}
\end{align}

\begin{align}
    P(\G)=\int_{\delta}P(\G|\delta)P(\delta)d\delta \label{eq:PG}
\end{align}

\noindent where $P(\text{Study success}|\delta)$ is the probability the trial is deemed a success based on the specified criteria for the trial (Bayesian or Frequentist) contingent on a particular value of $\delta$. If, for example, a standard frequentist success criterion is used where success is declared if $\hat{\delta}>c$ at significance level $\alpha$, then $c=\sigma_{\hat{\delta}}\Phi^{-1}(1-\alpha)+\delta^{\circ}$, where $\delta^{\circ}$ denotes the minimal clinically important efficacy and $\Phi^{-1}(x)$ is the inverse cumulative distribution function for the normal distribution. Using the semantic framework proposed by Best et al. \cite{Best2024}, we note $P_0(\delta)$ is a design prior as opposed to an analysis prior. The latter is used for the retrospective analysis of a study once data are in-hand, while the former is used to prospectively evaluate a study design across a range of potential true parameter values. Since the primary goal of a tripartite model is to model the relationship between decision tool and drug value in future studies, the posterior distribution of $\delta$ is given in terms of a design prior.

\vspace{0.2cm}

The assurance, or unconditional probability the trial is successful \cite{OHagan2005}, is given by

\begin{align}
    P(\text{Study success})=\int_{-\infty}^{\infty}P(\delta)d\delta \label{eq:suc}
\end{align}

\begin{figure}[H]
    \centering
    \includegraphics[scale=0.9]{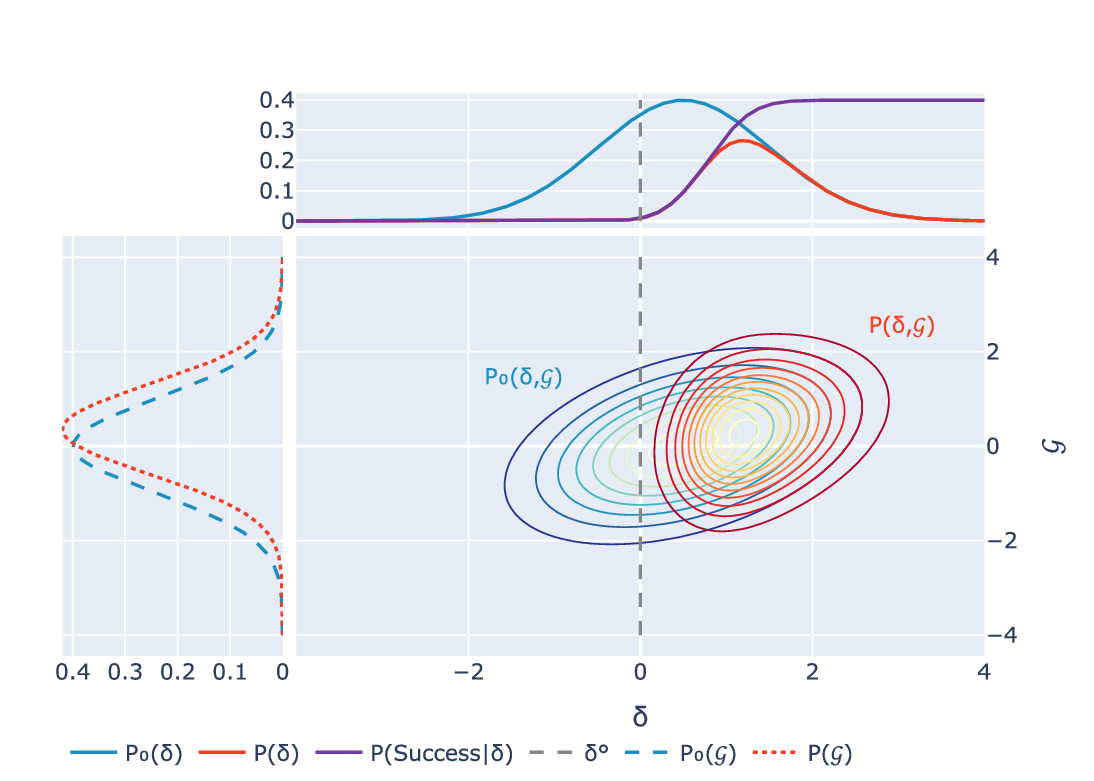}
    \caption{Tripartite model of a single stage clinical trial assuming a standard frequentist success criterion with $\delta^{\circ}=0$, $\alpha=0.025$, and $\sigma_{\hat{\delta}}=0.4$, thus $c=0.784$. $P_0(\delta)\sim\mathcal{N}(0.5,1.0)$ and $P_0(\G)\sim\mathcal{N}(0.0,1.0)$. The joint distribution of $\delta$ and $\G$ prior to the study, $P_0(\delta,\G)$, is assumed to be bivariate normal with $\rho=0.4$. Given $P_0(\delta,\G)$, if the trial is successful the new joint distribution of $\delta$ and $\G$ is denoted by $P(\delta,\G)$, and the marginal posteriors are $P(\delta)$ and $P(\G)$. Note that unlike $P_0(\delta,\G)$, $P(\delta,\G)$ is not bivariate normal. For visualization purposes $P(\text{Success}|\delta)$ has been scaled by the maximum of $P_0(\delta)$.}
    \label{fig:SingleModel}
\end{figure}

\noindent The probability the trial results in a False Positive, i.e. the trial is successful and $\delta<\delta^{\circ}$, is

\begin{align}
    P(\text{False positive}) = P(\delta<\delta^{\circ},\text{Study success})=\int_{-\infty}^{\delta^{\circ}}P(\delta)d\delta \label{eq:FP}
\end{align}

\noindent and the probability the trial results in a True Positive, i.e. the trial is successful and $\delta>\delta^{\circ}$, is

\begin{align}
    P(\text{True positive}) = P(\delta>\delta^{\circ},\text{Study success})=\int_{\delta^{\circ}}^{\infty}P(\delta)d\delta \label{eq:TP}
\end{align}

While the tripartite model is described using terms like ``study success” and ``trial”, words that typically denote clinical trials, this is simply because the language is more natural and it is important to recognize the model is equally applicable to preclinical settings where, instead of a clinical trial, one might run a virtual screen or in vitro screen. In these instances, the significance level $\alpha$ is typically not pre-specified or used at all, and instead the cutoff threshold $c$ is chosen directly, either as an absolute value (e.g., top five candidates advance) or as a relative value (e.g., top 0.1\% of candidates advance).

\vspace{0.2cm}

In the tripartite model the probability of a True Positive and False Positive are functions of the prior distribution of $\delta$, i.e. the estimate of the quality of the candidate drug(s) being evaluated, and the success criteria for the study, and are not uniquely determined by specifying the decision tool itself \cite{Brenner1997}. In practical terms this means that if two biopharma companies are using the same decision tool, the true and false positive rates and the quality of the candidates identified by the decision tool at the companies are not necessarily equivalent.

\begin{figure}[H]
    \centering
    \includegraphics[scale=0.9]{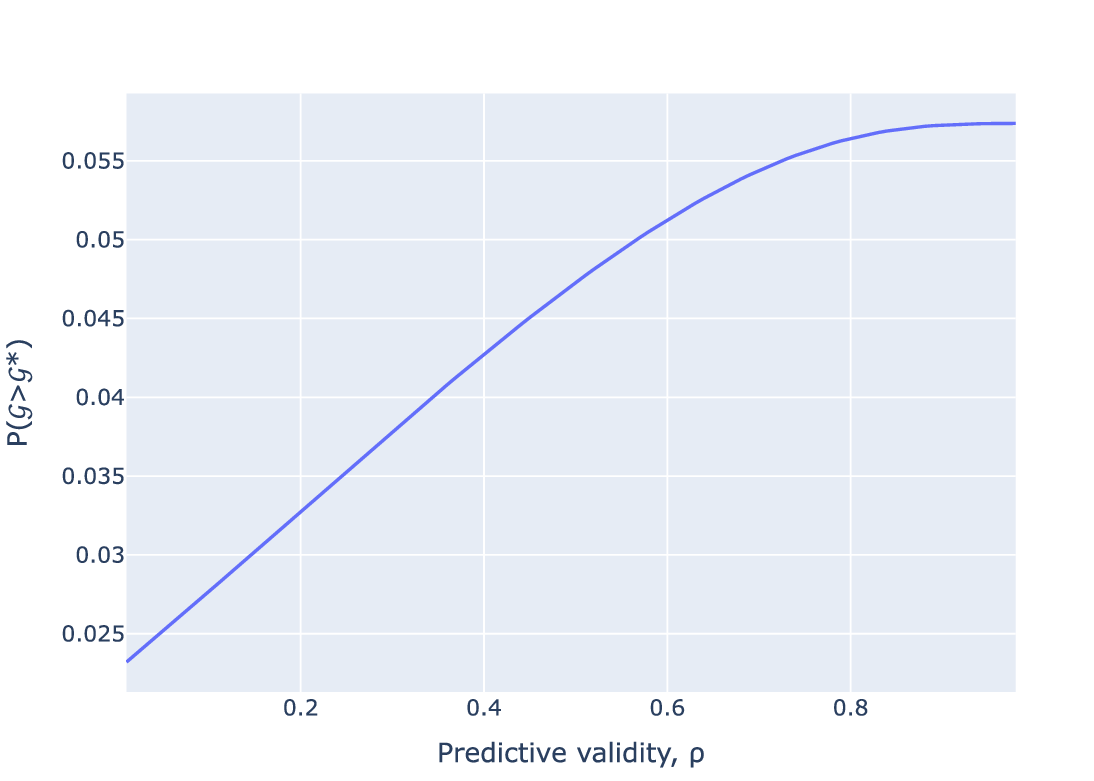}
    \caption{Probability that $\G>\G^*$ as a function of the predictive validity, $\rho$, of the decision tool for the tripartite model shown in Figure \ref{fig:SingleModel}.}
    \label{fig:SingleModelExt}
\end{figure}

\vspace{0.2cm}

Furthermore, we note that the probability of study success is not dependent on the relationship between $\delta$ and $\G$ (the predictive validity). This means that if two decision tools are used to evaluate the same drug candidate, the probability the candidate passes each test may be equivalent while the value of information gained is different. Importantly, probability of technical success should not be mistaken as an indicator of the quality of a study or value of the information gained during a study, and a more stringent test is not necessarily a better test.

\vspace{0.2cm}

Figure \ref{fig:SingleModel} depicts a tripartite model of a single-stage clinical trial (or preclinical screen) with a frequentist success criterion and illustrates a situation where there is modest initial optimism that the drug candidate will achieve the success criterion ($\delta>\delta^{\circ}$), indicated by $\mu_{\delta}=0.5$, in conjunction with initial ambivalence about the drug candidate’s overall potential for commercial and technical success, indicated by $\mu_{\G}=0$. The probability that the trial (or preclinical screen) is successful and, if successful, the resulting posterior distributions of $\delta$ and $\G$ are shown. 

\vspace{0.2cm}

If $\G^*$ is the threshold at which the drug generates a positive return versus alternative investments, the probability that $\G^*$ is exceeded given the study is successful can be easily calculated by integrating the marginal posterior $P(\G)$ from $\G^*$ to infinity. For the study modeled in Figure \ref{fig:SingleModel}, the probability that $\G$ exceeds $\G^*$ as a function of the study’s predictive validity when $\G^*=2$ is shown in Figure \ref{fig:SingleModelExt}. It is apparent that not all improvements in predictive validity are of equal value. An increase in predictive validity from 0.1 to 0.2 increases the probability of positive returns, i.e. $\G>\G^*$, by 18\%, while an increase in predictive validity from 0.8 to 0.9 increases the probability of positive returns by only 1.5\%. Crucially, these changes in the probability of positive returns are not dependent upon changes in the probability the study is successful\textemdash across all values of $\rho$ in Figure \ref{fig:SingleModelExt}, $P(\text{Study success})$ is constant at 39\%\textemdash but instead dependent on changes in the quality of the information gained by a successful study. 

\section{Tripartite Models in Series}
\label{sec:MultipleTripartite}

        Multiple tripartite models may be defined and placed in series with each other to describe a multi-stage drug development process (see also Scannell \& Bosley  for a decision-theoretic discussion of a multi-step R\&D process \cite{Scannell2016}). When placed in series, superscripts denote the stage to which the distribution belongs and all distributions are assumed to be posterior distributions unless indicated by a subscript zero, in which case they are prior distributions. Each stage has a predictive validity associated with it and the posterior marginal distribution of $\G$ at stage $n$, i.e., $P^n(\G)$, serves as the prior marginal distribution for $\G$ at stage $n+1$, i.e., $P_0^{n+1}(\G)$. In contrast, the posterior marginal of $\delta$ at stage $n$ need not be explicitly linked to the prior marginal of $\delta$ at stage $n+1$ and is thus free to be newly defined at each stage to account for changes in decision tools between the stages or more (or less) informative priors specific to the stage of development. 
        
        \begin{table}[H]
        \centering
\begin{tabular}{|c|llll|}
\hline
\multicolumn{1}{|l|}{\textbf{Parameter}} & \multicolumn{1}{l|}{\textbf{Stage 1}} & \multicolumn{1}{l|}{\textbf{Stage 2}} & \multicolumn{1}{l|}{\textbf{Stage 3}} & \textbf{Stage 4} \\ \hline
$\rho$                     & \multicolumn{1}{l|}{0.50}             & \multicolumn{1}{l|}{0.60}             & \multicolumn{1}{l|}{0.65}             & 0.80             \\ \hline
$\mu_{\delta}$                           & \multicolumn{1}{l|}{0.00}             & \multicolumn{1}{l|}{0.10}             & \multicolumn{1}{l|}{0.30}             & 0.80             \\ \hline
$\sigma_{\delta}$                        & \multicolumn{1}{l|}{0.90}             & \multicolumn{1}{l|}{0.55}             & \multicolumn{1}{l|}{1.50}             & 1.35             \\ \hline
$\sigma_{\hat{\delta}}$                  & \multicolumn{1}{l|}{1.05}             & \multicolumn{1}{l|}{0.30}             & \multicolumn{1}{l|}{0.60}             & 0.25             \\ \hline
$\delta^{\circ}$                         & \multicolumn{1}{l|}{0.10}             & \multicolumn{1}{l|}{0.00}             & \multicolumn{1}{l|}{0.00}             & 0.00             \\ \hline
$\alpha$                                 & \multicolumn{1}{l|}{0.77}             & \multicolumn{1}{l|}{0.45}             & \multicolumn{1}{l|}{0.05}             & 0.05             \\ \hline
$\mu_{\mathcal{G}}$                      & \multicolumn{4}{c|}{0.00}                                                                                                                \\ \hline
$\sigma_{\mathcal{G}}$                   & \multicolumn{4}{c|}{0.90}                                                                                                                \\ \hline
$\mathcal{G}^*$                          & \multicolumn{4}{c|}{1.49}                                                                                                                \\ \hline
\end{tabular}
\caption{Parameter values for a four-series tripartite model representing the preclinical, Phase 1, Phase 2, and Phase 3 in combination with regulatory approval stages of drug development. The value of $\G^*$ was chosen such that the initial drug candidate has an approximately 5\% chance of generating a positive return.}
    \label{tab:Params}
\end{table}
        
        \vspace{0.2cm}
        
\begin{figure}[H]
    \centering
    \includegraphics[scale=0.9]{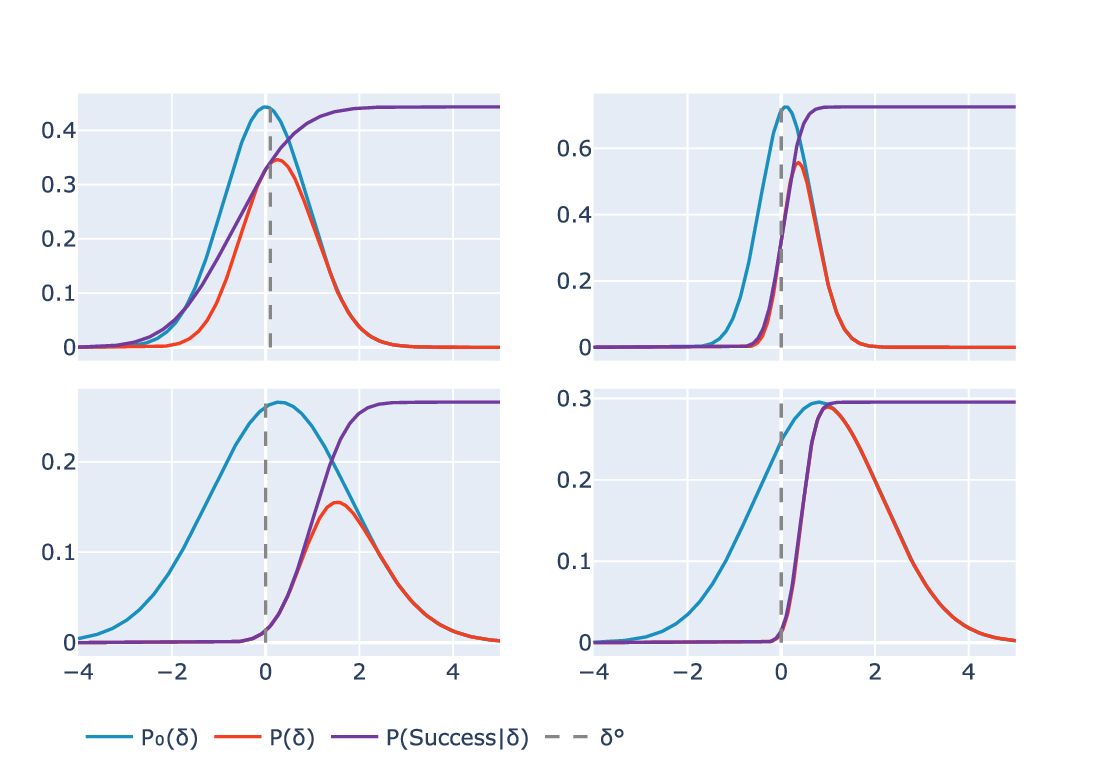}
    \caption{Trial-related parameters for a four-series tripartite model parameterized by Table \ref{tab:Params} representing the preclinical (top left), Phase 1 (top right), Phase 2 (bottom left), and Phase 3 and regulatory (bottom right) stages of drug development. For visualization purposes $P(\text{Success}|\delta)$ has been scaled by the maximum of $P_0(\delta)$.}
    \label{fig:FourSeriesa}
\end{figure}

\begin{figure}[H]
    \centering
    \includegraphics[scale=0.9]{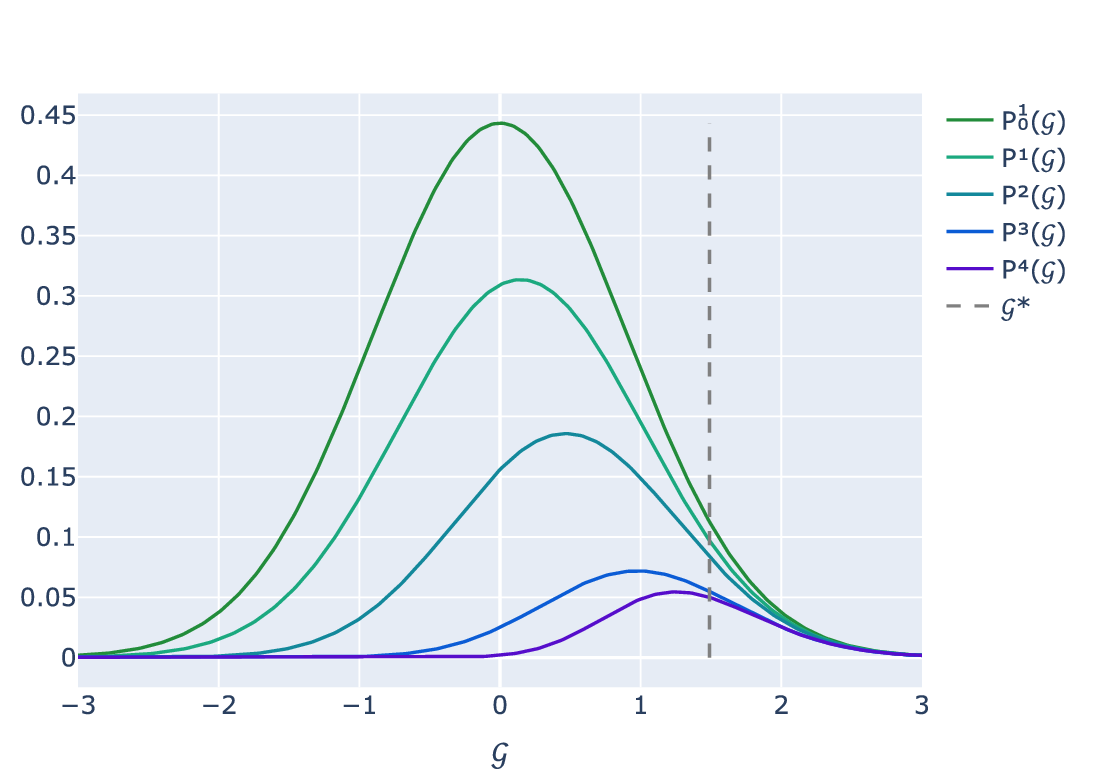}
    \caption{Probability distributions of $\G$ for the four-series tripartite model parameterized by Table \ref{tab:Params}.}
    \label{fig:FourSeriesb}
\end{figure}

\begin{figure}[H]
    \centering
    \includegraphics[scale=0.9]{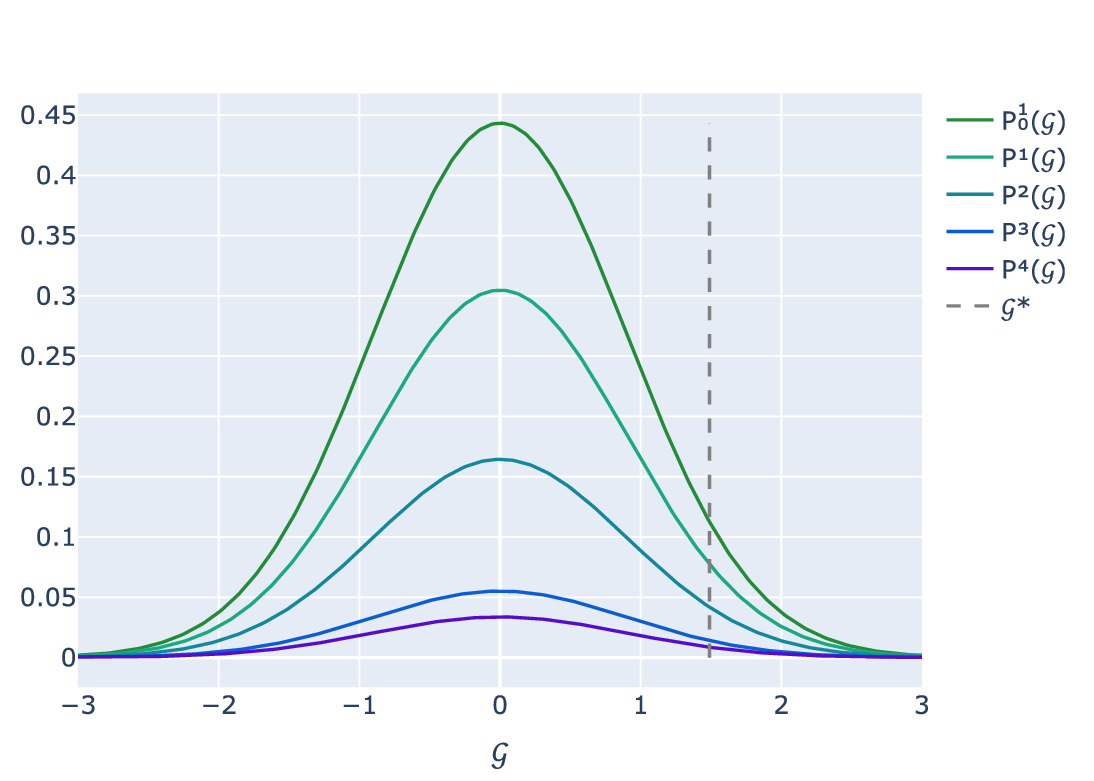}
    \caption{Four-series of tripartite models with $\rho=0$ at each stage. All other parameters are the same as in Table \ref{tab:Params}.}
    \label{fig:FourSeriesc}
\end{figure}

        Table \ref{tab:Params} shows parameter values used in a four-series tripartite model—that is, four tripartite models arranged in series—corresponding to the preclinical, Phase 1, Phase 2 and, in combination, Phase 3 and regulatory stages of the drug development process. The parameter values were chosen to approximate the stepwise progression probabilities reported by Paul et al. \cite{Paul2010} and to match industry-standard values to the extent they exist, e.g. $\alpha=0.05$ in Phase 2 and 3.  The resulting model and summary statistics of the model are shown in Figures \ref{fig:FourSeriesa} and \ref{fig:FourSeriesb}, and Table \ref{tab:Output}, respectively. The total area under the marginal posteriors of $\G$ decreases with the stage of development, indicating the decreasing likelihood the drug candidate will make it to later stages without being eliminated. Simultaneously, as information is gained about the drug candidate during successive development stages, an increasing proportion of the area under the marginal posterior of $\G$ exceeds the threshold $\G^*$, indicating that the estimate of the drug’s capacity for commercial and technical success is increasing as it progresses to later development stages. The value of the drug candidate increases through the development process due to both the progressive reduction in risk the drug will fail a development stage and thus never reach the market, and the positive predictive validity of the decision tools used at each stage of development. If instead, the decision tools used at each stage had zero predictive validity, the probability that $\G$ exceeds $\G^*$ as the drug candidate advances toward approval remains constant, signifying that no information is being gained about the drug candidate’s capacity for commercial and technical success even as it passes successive stages of development (Figure \ref{fig:FourSeriesc}).
        
\begin{table}[H]
\centering
\begin{tabular}{|c|l|l|l|l|}
\hline
                                                 & \textbf{Stage 1} & \textbf{Stage 2} & \textbf{Stage 3} & \textbf{Stage 4} \\ \hline
$P(\text{Study success})$                        & 0.687            & 0.540            & 0.335            & 0.611            \\ \hline
Cumulative $P(\text{Study success})$             & 0.687            & 0.371            & 0.124            & 0.076            \\ \hline
$P(\text{Study success}, \delta>\delta^{\circ})$ & 0.408            & 0.476            & 0.332            & 0.610            \\ \hline
$P(\text{Study success},\delta<\delta^{\circ})$  & 0.279            & 0.063            & 0.003            & 0.001            \\ \hline
$P(\delta>\delta^{\circ}|\text{Study success})$  & 0.593            & 0.882            & 0.991            & 0.998            \\ \hline
$P(\delta<\delta^{\circ}|\text{Study success})$  & 0.407            & 0.118            & 0.009            & 0.002            \\ \hline
$P(\G>\G^*|\text{Study success})$                & 0.063            & 0.106            & 0.242            & 0.384            \\ \hline
\end{tabular}
\caption{Descriptive statistics of the four-series tripartite model shown in Figures \ref{fig:FourSeriesa} and \ref{fig:FourSeriesb}, and parameterized with value in Table \ref{tab:Params}. The value of $P(\G>\G^*|\text{Study success})$ at the completion of Stage 4, in this case 0.384, is the terminal probability that $\G$ exceeds $\G^*$ assuming all development stages were successful.}
    \label{tab:Output}
\end{table}
        
\vspace{0.2cm}

By placing tripartite models in series, we can also determine how changes in the predictive validity of decision tools used in early stages of drug development influence the value of later stages and eventually inform the terminal value of the drug in a formal manner. To do this, we simply note the change in the terminal probability that $\G$ exceeds the threshold $\G^*$ after Stage 4, i.e. $P^4(G>\G^*|\text{Study success})$, and multiply this by the expected market value of the drug when $\G>\G^*$.
\noindent If a decision tool is replaced with a new decision tool that increases $P^4(G>\G^*|\text{Study success})$ by 1 percentage point, e.g. from 3\% to 4\%, and the market value of the drug candidate is \$100, then the value provided by the new decision tool is \$1. In this way the financial value of the decision tool is a function of both the quality of the decision tool and the market value of the drug candidate, and one cannot estimate the financial value of a decision tool without specifying the value of the drug candidate’s target market.

\vspace{0.2cm}

\begin{figure}[H]
    \centering
    \includegraphics[scale=0.9]{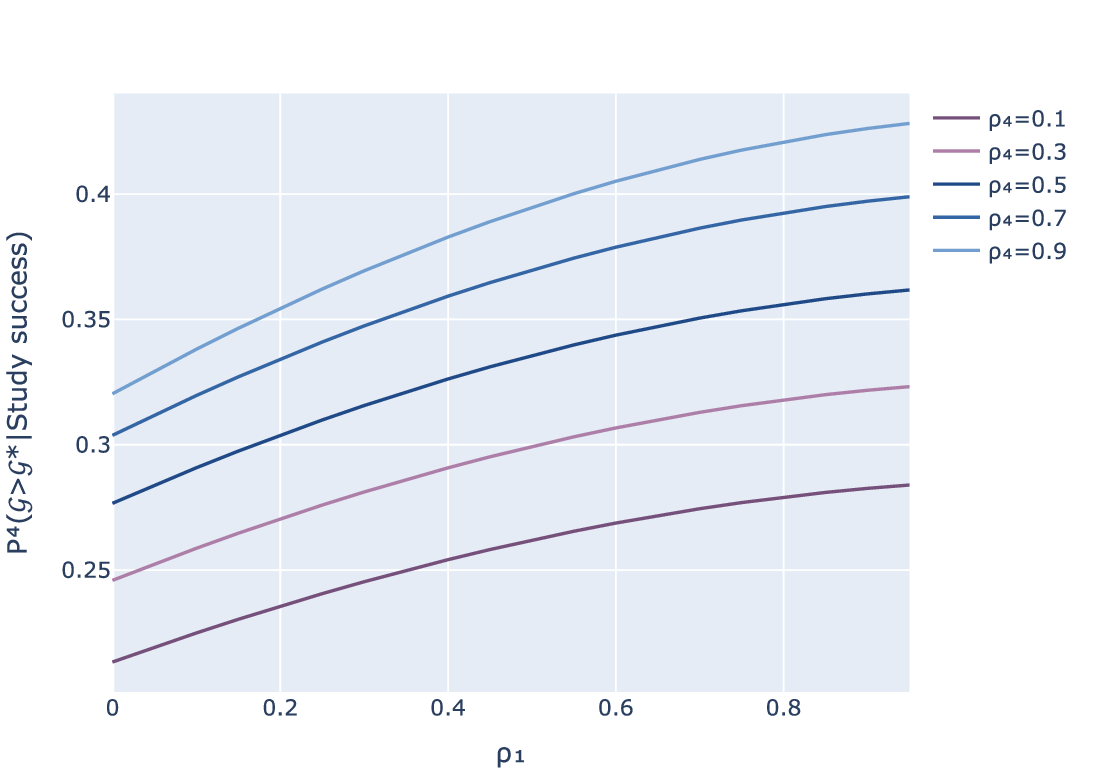}
    \caption{Changes in the terminal probability that $\G$ exceeds $\G^*$ as a function of changes in the predictive validity of the preclinical ($\rho_1$) and Phase 3 and regulatory ($\rho_4$) stages of development.}
    \label{fig:FourSeriesd}
\end{figure}

We can further investigate the interplay between the predictive validity of decision tools at various stages and their collective impact on $P^4(G>\G^*|\text{Study success})$. The four-series tripartite model illustrates that the value of increases in the predictive validity of decision tools at earlier stages of development can be limited by the predictive validity of decision tools at later stages. Figure \ref{fig:FourSeriesd} shows $P^4(G>\G^*|\text{Study success})$ as a function of the predictive validity at Stage 1 ($\rho_1$) for various values of the predictive validity at Stage 3 ($\rho_3$). When $\rho_3=0.1$, an increase in $\rho_1$ from 0.2 to 0.4 increases $P^4(G>\G^*|\text{Study success})$  by 7.6\%, whereas when $\rho_3=0.9$, increasing $\rho_1$ from 0.2 to 0.4 increases $P^4(G>\G^*|\text{Study success})$  by 8.6\%.

\subsection{Discussion and Conclusions}
\label{subsec:Discussion}

        There are arguments in the economics and pharmaceutical R\&D literatures \cite{Scannell2022,BillettedeVillemeur2019, BillettedeVillemeur2022} that – from the perspective of therapeutic progress – too much capital flows into the creation of new therapeutic candidates and too little capital flows into the creation of new decision tools that better distinguish between good and bad candidates. 

\vspace{0.2cm}

Such relative over-investment in therapeutic candidates and relative under-investment in decision tools is likely to be important in practical terms because decision tools’ predictive validity is a major constraint on biopharmaceutical R\&D productivity \cite{Scannell2016, Scannell2022}. Furthermore, decision tools are an economic complement of therapeutic candidates, and vice-versa.

\vspace{0.2cm}

The existing literature places blame primarily on variance in the economic appropriability of different kinds of knowledge. Therapeutic candidates are generally protected by patents and other forms of market exclusivity that allow inventors to capture much of the economic value of their invention. New decision tools, on the other hand, tend to create knowledge that spills-over, which benefits other players who have not invested in creating the new decision tools. To paraphrase this literature, private sector players see better returns from playing low probability chemical roulette than from creating new tools that improve the odds of the game.  

\vspace{0.2cm}

The contrast between the structure of the tripartite model in this paper, and the conventional rNPV approach, points to another cause of under-investment in decision tools: \emph{model specification bias}. Investors who rely on conventional rNPV methods find it relatively easy to assign a value to therapeutic candidates but have no practical way of assigning value to the decision tools that determine the candidate’s correct or incorrect progression. Since they can’t assign value to the decision tools themselves, or assess the tools’ impact on the value of drug candidates, they fail to allocate sufficient capital to improve decision tool performance.

\vspace{0.2cm} 

In contrast to rNPV models, the multi-stage tripartite approach we present here can simultaneously value therapeutic candidates, individual decision steps, and important interactions within the R\&D process. These interactions include the downstream consequences of better or worse upstream decisions, how the value of a decision tool, and of decision tool improvement, varies with the size of the drug market, and with the performance of upstream or downstream decision tools. The tripartite approach may therefore support more effective allocation of R\&D capital. It may support investment in therapeutic assets that are developed using good decision tools, and also direct investment towards better decision tools which will find more good therapeutic assets. 

\bibliographystyle{unsrt}  
\bibliography{Refs}

\end{document}